\documentclass[preprint, superscriptaddress]{revtex4-1}
\usepackage{graphicx}
\usepackage{dcolumn}
\usepackage{amsmath}
\usepackage{amsfonts}
\usepackage{mathrsfs}
\usepackage{amssymb}
\usepackage{physics}
\usepackage{mathtools}
\usepackage{stmaryrd} 
\usepackage{xcolor}
\usepackage{tcolorbox}
\usepackage{orcidlink}

\usepackage{algorithm}
\usepackage{algpseudocode}
\usepackage{hyperref}
\hypersetup{
	colorlinks=true,
	linkcolor=blue,
	filecolor=blue,      
	urlcolor=blue,
	citecolor=blue,
	pdfstartview={XYZ null null 1.07}
}

\usepackage[T1]{fontenc}
\usepackage[latin1]{inputenc}
\usepackage{booktabs}
\usepackage[font=small,labelfont=bf,tableposition=top]{caption}
\usepackage{rotating}
\usepackage{array,multirow}
\usepackage{float}

\usepackage{makecell} 

\begin{document}
\begin{center}
\hrule height 4.1pt
\vspace{0.5cm}
{\Large\textbf{Electronic Structure Calculations using Quantum Computing}}
\vspace{0.5cm}
\hrule height 1.2pt
\end{center}

\author{Nouhaila Innan\orcidlink{0000-0002-1014-3457}}\email[]{nouhailainnan@gmail.com}
\affiliation{\footnotesize Quantum Physics and Magnetism Team, LPMC, Faculty of Sciences Ben M'sick, Hassan II University of Casablanca, Morocco}
\affiliation{\footnotesize Quantum Formalism Fellow, Zaiku Group Ltd, Liverpool, United Kingdom.}

\author{Muhammad Al-Zafar Khan\orcidlink{0000-0002-1147-7782}}\email[]{muhammadalzafark@gmail.com}
\affiliation{\footnotesize Quantum Formalism Fellow, Zaiku Group Ltd, Liverpool, United Kingdom.}
\affiliation{\footnotesize Robotics, Autonomous Intelligence, and Learning Laboratory (RAIL), School of Computer Science and Applied Mathematics, University of the Witwatersrand, 1 Jan Smuts Ave, Braamfontein, Johannesburg 2000, Gauteng, South Africa}

\author{Mohamed Bennai\orcidlink{0000-0002-7364-5171}}
\affiliation{\footnotesize Quantum Physics and Magnetism Team, LPMC, Faculty of Sciences Ben M'sick, Hassan II University of Casablanca, Morocco}

\begin{abstract}
The computation of electronic structure properties at the quantum level is a crucial aspect of modern physics research. However, conventional methods can be computationally demanding for larger, more complex systems. To address this issue, we present a hybrid Classical-Quantum computational procedure that uses the Variational Quantum Eigensolver (VQE) algorithm. By mapping the quantum system to a set of qubits and utilising a quantum circuit to prepare the ground state wavefunction, our algorithm offers a streamlined process requiring fewer computational resources than classical methods. Our algorithm demonstrated similar accuracy in rigorous comparisons with conventional electronic structure methods, such as Density Functional Theory and Hartree-Fock Theory, on a range of molecules while utilising significantly fewer resources. These results indicate the potential of the algorithm to expedite the development of new materials and technologies. This work paves the way for overcoming the computational challenges of electronic structure calculations. It demonstrates the transformative impact of quantum computing on advancing our understanding of complex quantum systems. 
\end{abstract}

\maketitle


\section{Introduction}
\textit{Atoms} are the constituent building blocks of all organic and inorganic organisms. Many atoms group together to form \text{molecules}, and arrangements of these molecules in certain configurations give rise to the complex diversity of nature and, by extension, to inanimate objects. 

Electronic structure calculations are used to determine the properties and behaviours of these atoms and molecules. Since these calculations are performed at the atomic scales, an intrinsic paradigm and framework for determining these properties is quantum mechanics, specifically via the generalised, time-dependent, Schr\"{o}dinger Eq. \eqref{eq:schordinger}:
\begin{equation}
\label{eq:schordinger}
\imath\hbar\frac{\partial\Psi(\mathbf{r},t)}{\partial t}=-\frac{\hbar^{2}}{2m}\boldsymbol{\nabla}\cdot\boldsymbol{\nabla}\Psi(\mathbf{r},t)+V(\mathbf{r},t)\Psi(\mathbf{r},t)\Longleftrightarrow \hat{E}\ket{\psi(\mathbf{r},t)}=\hat{H}\ket{\Psi(\mathbf{r},t)},
\end{equation}
where $\Psi(\mathbf{r},t)$ is the wavefunction that is dependent on three-dimensional space $\mathbf{r}$ and time $t$, $V(\mathbf{r},t)$ is the potential energy function, $\hbar\approx 1.054\times 10^{-34}\;\text{J}.\text{s}$ is the reduced Planck constant, $m$ is the particle mass, and $\imath=\sqrt{-1}$. The equivalency relation renders the generalised Schr\"{o}dinger equation in terms of the energy operator: $\hat{E}\square\overset{\Delta}{=}\imath\hbar\frac{\partial\square}{\partial t}$, and the Hamiltonian operator: $\hat{H}\square\overset{\Delta}{=}\left[-\frac{\hbar^{2}}{2m}\nabla^{2}+V(\mathbf{r},t)\right]\square$. 

The Schr\"{o}dinger equation is a nonlinear partial differential equation (PDE) that does not have an explicit solution for general potential energies. Thus, the need to approximate this solution has arisen. Within the context of electronic structure calculations, the most famous ``self-consistent'' field method introduced by \hyperlink{1}{Hartree, 1928} and by \hyperlink{2}{Fock, 1930} known famously as ``Hartree-Fock theory'' (HF), is used to approximate the Schr\"{o}dinger equation for many-electron systems via Eq. \eqref{eq:hf}:
\begin{equation}
\label{eq:hf}
\left(\mathcal{F}-\epsilon_{i}\right)\Psi_{i}(\mathbf{r})=0,\quad \mathcal{F}\square=\left[-\frac{1}{2}\nabla^{2}+V(\mathbf{r})+\mathcal{J}(\mathbf{r})-\mathcal{K}(\mathbf{r})\right]\square,\quad \mathcal{J}(\mathbf{r})=\int_{\Omega}\frac{\rho(\mathbf{r}')}{|\mathbf{r}-\mathbf{r}'|}\;d^{3}\mathbf{r}'.
\end{equation}
where $\mathcal{F}$ is the Fock operator consisting of the external potential energy, $V(\mathbf{r})$, of nuclei; the Coulomb operator, $\mathcal{J}(\mathbf{r})$, which describes the interaction between electrons over the volume distribution $\Omega$ with charge distribution $\rho$; $\mathcal{K}(\mathbf{r})$ is the exchange operator which accounts for the antisymmetry of the electron wavefunction; $\epsilon_{i}$ is the energy of the $i^{\text{th}}$ electron; and $\Psi_{i}(\mathbf{r})$ is the molecular orbital of the $i^{\text{th}}$ electron. Typically, the Fock operator is expressed as the density matrix in Eq. \eqref{eq:density}: 
\begin{equation}
\label{eq:density}
P=\sum_{i=1}^{n\;\text{electrons}}\Psi_{i}(\mathbf{r})\Psi_{i}^{*}(\mathbf{r}).
\end{equation}

Density Functional Theory (DFT), introduced by \hyperlink{3}{Hohenberg \& Kohn, 1964}, makes use of the Kohn-Sham equations to determine the electronic structure of molecules and works exceptionally well in solids. By theoretically demonstrating via the Hohenberg-Kohn theorem that the ground state electron density of a system is a unique functional of the external potential, and vice versa, the total energy of the system, and all other properties, can be inferred. Mathematically, these ideas are encapsulated in Eq. \eqref{eq:kohnsham}:
\begin{equation}
\label{eq:kohnsham}
\left[-\frac{1}{2}\nabla^{2}+V(\mathbf{r})\right]\rho(\mathbf{r})=\epsilon_{i}\rho(\mathbf{r}),
\end{equation}
where $\rho(\mathbf{r})$ serves as the electron density, $V(\mathbf{r})$ is the effective potential, and $\epsilon_{i}$ is the energy of the $i^{\text{th}}$ electron. 

Modern DFT builds upon the work of \hyperlink{3}{Hohenberg \& Kohn, 1964} and has become a fundamental tool in the arsenal of the contemporary Computational Chemist and Materials Scientist when determining the electronic structure of molecules and solids. 

Many-body Perturbation Theory (MBPT), introduced by \hyperlink{4}{Luttinger \& Ward, 1960}, and expanded upon by \hyperlink{6}{Baym, 1962} and \hyperlink{5}{Hedin, 1965}, built upon the work of \hyperlink{7}{Schwinger, 1951} and \hyperlink{8}{Matsubara, 1955}, combines perturbation theory with Green's functions to calculate the electronic structure of many-electron systems. Mathematically, the ideas of MBPT are encapsulated by the Dyson equation (\hyperlink{9}{Dyson, 1949}) as expressed in Eq. \eqref{eq:dyson}:
\begin{equation}
\label{eq:dyson}
G(\mathbf{r}',\mathbf{r})=G_{0}(\mathbf{r}',\mathbf{r})+G_{0}(\mathbf{r}',\mathbf{r})\Sigma G(\mathbf{r}',\mathbf{r}),
\end{equation}
where $G_{0}$ is the non-interacting Green's function, $G$ is the full Green's function, and $\Sigma$ is the self-energy -- which serves as a summation over all particle-hole excitations expressed as Eq. \eqref{eq:hole}:
\begin{equation}
\label{eq:hole}
\Sigma=-V\int d^{3}\mathbf{r}\int d^{3}\mathbf{r}'v(|\mathbf{r}-\mathbf{r}'|)\mathcal{X}(\mathbf{r},\mathbf{r}')G_{0}(\mathbf{r}',\mathbf{r}),\quad\quad \mathcal{X}(\mathbf{r},\mathbf{r}')=-\frac{1}{V}\frac{\partial^{2}E}{\partial F^{2}}, 
\end{equation}
where $V$ is the system volume, $E$ is the total energy of the system, $v(|\mathbf{r}-\mathbf{r}'|)$ is the Coulomb interaction potential between particles, $\mathcal{X}(\mathbf{r},\mathbf{r}')$ is the rank-2 polarisability tensor that relates the induced dipole moment to the strength and direction of the external electric field, and $F$ is the strength of the external electric field. 

Coupled Cluster (CC) is an advancement of the Hartree-Fock method, described above, introduced by \hyperlink{10}{Coester \& K\"{u}mmel, 1960}, serves as a highly accurate method for computing molecular properties. Initially developed for calculating nuclear binding energies and several other properties, the technique is now used to model the electronic wavefunction of solid-state systems, precisely, transition metal complexes, bond dissociation energies, excitation energies, dipole moments, potential energy surfaces, and reaction barriers. Mathematically, the technique is expressed in terms of the exponential approximation as Eq. \eqref{eq:exponn}:
\begin{equation}
\label{eq:exponn}
\ket{\Psi_{CC}}=e^{\mathcal{T}}\ket{\Phi_{0}},
\end{equation} 
where $\ket{\Psi_{CC}}$ is the CC wavefunction, $\ket{\Phi_{0}}$ is the Hartree-Fock reference state which defines single-particle orbitals and occupation numbers, and $\mathcal{T}$ is the cluster operator expressed as Eq. \eqref{eq:Tcluster}:
\begin{equation}
\label{eq:Tcluster}
\mathcal{T}=\sum_{i}T_{i},
\end{equation}
where $T_{i}$ is the $i^{\text{th}}$-fold excitation and the summation runs over all excitations.

Configuration Interactions (CI) is conceptually the simplest method for solving the time-independent Schr\"{o}dinger equation under the Born-Oppenheimer approximation (\hyperlink{18}{Sherrill \& Schaefer III, 1999}). The method constructs a trial wavefunction as a linear combination of Slater determinants that correspond to different electron configurations of the system under examination. Using combinatorics, the Slater determinants are constructed by apportioning the electrons to the available orbitals. The CI method is broken up into two sub-methods: The full CI (FCI) method, which considers all possible configurations, and the truncated CI (TCI) method, which considers subsets of the configurations. The wavefunction for the CI method is constructed using Eq. \eqref{eq:tci}:
\begin{equation}
\label{eq:tci}
\Psi=\sum_{i=0}^{n}c_{i}\phi_{i},
\end{equation}
where $c_{i}$ are coefficients of the Slater determinants $\phi_{i}$, for $0\leq i\leq n$. By solving a system of linear equations, the coefficients can be obtained. Once the coefficients are determined, the total energy can be calculated using Eq. \eqref{eq:slater}:

\begin{equation}
\label{eq:slater}
\mathcal{E}=\frac{\sum_{i=0}^{n}c_{i}^{2}E_{i}}{\sum_{i=0}^{n}c_{i}^{2}},
\end{equation}
where $E_{i}$ are the energies corresponding to the Slater determinants $\phi_{i}$. 

Time-dependent DFT (TDDFT), introduced by \hyperlink{19}{Runge \& Gross, 1984}, is an extension of regular DFT to calculate the ground state electronic structure of a temporally evolving system. In this approach, the wavefunction, comprising the electron density of external potential, is described by the Kohn-Sham equations' time-dependent, coupled PDE system. This approach is widely adopted in the studies of theoretical spectroscopy (\hyperlink{21}{Besley \& Asmuruf, 2010}), photochemistry (\hyperlink{20}{Matsuzawa \textit{et al.}, 2001}), and energy transference. The goal is to approximate the solutions to the time-dependent Schr\"{o}dinger equation \eqref{eq:schordinger} above. Analogously, the time-dependent DFT equation is given by Eq. \eqref{eq:dfteq}:
\begin{equation}
\label{eq:dfteq}
\hat{E}\rho(\mathbf{r},t)=\left[\hat{H},\rho(\mathbf{r},t)\right],
\end{equation}
where $\rho(\mathbf{r},t)$ is the time-dependent electron density. The corresponding time-dependent Kohn-Sham equation is expressed as Eq. \eqref{eq:khtime}:
\begin{equation}
\label{eq:khtime}
\hat{E}\Psi_{i}(\mathbf{r},t)=\left[\hat{H}_{KS}(t)+\hat{\Sigma}(\mathbf{r},t)\right]\Psi_{i}(\mathbf{r},t).
\end{equation}
where $\hat{H}_{\text{KS}}(t)$ is the time-dependent Kohn-Sham Hamiltonian given by Eq. \eqref{eq:hkst}:
\begin{align}
\label{eq:hkst}
\hat{H}_{KS}(t)=&\;-\frac{1}{2}\sum_{i=1}^{n}\nabla_{i}^{2}+\int\frac{\tilde{\rho}(\mathbf{r},t)}{\mathbf{r}-\mathbf{r}'}\;d\mathbf{r}'+\int v_{\text{ext}}(\mathbf{r},t)\tilde{\rho}(\mathbf{r},t)\;d\mathbf{r} \nonumber \\
&\;+\int v_{\text{Hxc}}(\mathbf{r},t)\rho(\mathbf{r},t)\;d\mathbf{r}+\int v_{\text{xc}}\delta\tilde{\rho}(\mathbf{r},t)\;d\mathbf{r},
\end{align}
where $v_{\text{ext}}(\mathbf{r},t)$ is the time-dependent external potential, $v_{\text{Hxc}}(\mathbf{r},t)$ is the time-dependent Hartree potential, $v_{\text{xc}}(\mathbf{r},t)$ is the time-dependent exchange-correlation potential, and $\delta\tilde{\rho}(\mathbf{r},t)$ is the deviation of the energy density from the ground state. The exchange-correlation self-energy operator, $\hat{\Sigma}(\mathbf{r},t)$ is given by Eq. \eqref{eq:groundenrg}:
\begin{equation}
\label{eq:groundenrg}
\hat{\Sigma}(\mathbf{r},t)=\int d^{3}\mathbf{r}'\int dt'\frac{\delta E_{\text{xc}}[\tilde{\rho}]}{\delta\tilde{\rho}(\mathbf{r}',t')}\delta(\mathbf{r}-\mathbf{r}')\delta(t-t'),
\end{equation} 
where $E_{\text{xc}}[\tilde{\rho}]$ is the exchange-correlation energy functional; $\frac{\delta E_{\text{xc}}[\tilde{\rho}]}{\delta\tilde{\rho}(\mathbf{r}',t')}$ is the functional derivative of the exchange-correlation energy with respect to the energy density;  $\delta(\mathbf{r}-\mathbf{r}')$, and $\delta(t-t')$ are the spatial and temporal Dirac delta functions. 

As one may gauge, TDDFT is highly accurate in describing the ground state energies of many-electron systems; it is mathematically laborious to implement and computationally expensive. Thus, the impetus for an easier technique. 

Quantum Monte Carlo (QMC), see \hyperlink{22}{Foulkes \textit{et al.}, 2001}; \hyperlink{23}{Frank \textit{et al.}, 2019}, is a statistical precude used to sample a many-body wavefunction of a system. At the heart of the method is the generation of a large number of random samples generated from the wavefunction. There are several QMC methods used for the simulation of molecules. These include: The Metropolis algorithm (\hyperlink{24}{Metropolis \textit{et al.}, 1953}), Variational MC (VMC) -- see \hyperlink{25}{Kalos, 1962}, Diffusion MC (DMC) -- see \hyperlink{26}{Barnett \& Whaley, 1993}, Green's Function MC (GFMC) -- see \hyperlink{27}{Trivedi \& Ceperley, 1990}, Auxiliary Field Quantum MC (AFQMC) -- see \hyperlink{28}{Lee \textit{et al.}, 2022}, Projector Quantum MC (PQMC) -- see \hyperlink{29}{Hetzel \textit{et al.}, 1997}, Path Integral MC (PIMC) -- see \hyperlink{30}{Barker, 1979} and \hyperlink{31}{Cazorla \& Boronat, 2017}, Stochastic Reconfiguration -- see \hyperlink{32}{Sorella, 1998}, and Population Annealing MC (PAMC) -- see \hyperlink{33}{Weigel \textit{et al.}, 2021}, amongst others. Below, we present the simplest of these methods, the Metropolis algorithm \ref{metropolis}.
\begin{algorithm}[H]
\caption{\texttt{Metropolis}$(s, \tilde{s}, E(s), E(\tilde{s}), T)$}
\begin{algorithmic}
\State \texttt{// the algorithm takes in a state $s$, candidate state $\tilde{s}$, corresponding energies $E(s)$ and $E(\tilde{s})$ respectively, and temperature $T$} 
\State \textbf{input} the number of samples $n$ 
\Repeat 
\State calculate the change in energy $\Delta E=E(\tilde{s})-E(s)$ 
\If{$\Delta E<0$} 
\State accept candidate state $\tilde{s}$ 
\State set $P\longleftarrow 1$ \texttt{// set the probability to 1} 
\State set $s\longleftarrow s'$ 
\ElsIf{$\Delta E>0$ }
\State accept candidate state $\tilde{s}$ 
\State set $P\longleftarrow \exp\left(-\frac{\Delta E}{T}\right)$ \texttt{// set the probability to a decaying exponential of the ratio of the energy difference and temperature} 
\State set $s\longleftarrow s'$ 
\Else
\State \textbf{pass} 
\EndIf
\Until{$n$ samples are obtained} 
\State \textbf{return} a set of configurations of the system being studied 
\end{algorithmic}
\label{metropolis}
\end{algorithm}
\vspace{-0.5cm}
Molecular Dynamics (MD), introduced by the landmark paper by \hyperlink{41}{Alder \& Wainwright, 1959}, and expanded upon by the Nobel Laureate, Martin Karplus (1930-), and his research group. The aim of MD is to use the laws of Classical and Quantum Mechanics to study the motion of molecules over time. Since the mathematical equations are well known, it will be an exercise in futility to restate them here; see the excellent source by \hyperlink{42}{Smit \& Frenkel, 1996} for a detailed discussion.

Some of the important features that these methods can be used to calculate are tabulated below (\ref{tab3} \& \ref{tab4}).
\begin{table}[H] 
\caption{Uses of the various electronic structure calculation methods, and computational complexities -- Part I.}

\centering
\begin{tabular}{|p{5cm}|p{5cm}|p{7cm}|}
\hline
\hline
\textbf{Method} &\textbf{Uses} &\textbf{Computational Complexity} \\
\hline
Hartree-Fock Theory &
 $\bullet$  Electron correlation effects. \newline
 $\bullet$  Closed-shell systems. \newline
 $\bullet$  Ground state electronic structure of molecules and solids. &
$\mathcal{O}(N^{4})$, where $N$ is the number of basis functions used to expand the wavefunction. \\
\hline 
Density Functional Theory &$\bullet$ Systems with a large number of atoms. &$\mathcal{O}(N^{3})$, where $N$ is the number of atoms in the system. \\
 &$\bullet$ Systems with a large number of electrons. & \\
 &$\bullet$ Used to calculate electronic structure, stability, and reactivity. & \\
\hline 
Many-Body Perturbation \newline Theory &
$\bullet$ Used extensively in condensed matter physics. \newline
$\bullet$ Used to describe materials with strong electron-electron correlations, such as rare earth metals, transition metal oxides, and heavy fermion materials. \newline
$\bullet$ Used to describe small systems.
& 
$\mathcal{O}(N^{4})-\mathcal{O}(N^{6})$, where $N$ is the number of atoms in the system / basis functions used to describe the system. For second-order perturbations: $\mathcal{O}(N^{5})$, for third-order perturbations: $\mathcal{O}(N^{6})$, etc. \\
\hline 
\end{tabular}
\label{tab3}
\end{table}

\begin{table}[H]
\caption{Uses of the various electronic structure calculation methods, and computational complexities -- Part II.}
\centering
\begin{tabular}{|p{5cm}|p{5cm}|p{7cm}|}
\hline
\hline
\textbf{Method} &\textbf{Uses} &\textbf{Computational Complexity} \\
\hline 
& $\bullet$ Used to describe specific regions of large systems. & \\
\hline
Coupled Cluster &$\bullet$ Considered the gold standard for predicting molecular properties. &$\mathcal{O}(N^{6})$ for the Coupled Cluster Singles and Doubles (CCSD) method, where $N$ is the number of molecular orbitals. \\
 &$\bullet$ Used to predict bond breaking, reaction pathways, and excited state energies. &The exponential computational complexity can be reduced by employing reduced Coupled Cluster methods or by parallelising the calculations. \\  
\hline 
Configuration Interaction &$\bullet$ Well-suited for systems with strong electron correlation. &$\mathcal{O}(N^{k}\times n_{e^{-}}^{m}\times n_{v}^{p})$, where $N$ is the number of spin-orbitals, $n_{e^{-}}$ is the number of electrons, $n_{v}$ is the number of virtual orbitals, $m$ is the number of singly-excited determinants in the calculation, and $p$ is the number of doubly-excited determinants in the calculation. \\
\hline 
Time-Dependent Density \newline Functional Theory &$\bullet$ Used to study electronic excitations, chemical reactions, and charge transfer processes. &$\mathcal{O}(N^{3})$, where $N$ is the number of electrons in the system. \\
 &$\bullet$ Used to study phonons, excitons, and plasmons. & \\
 &$\bullet$ Used to calculate transition dipoles, Rydberg states, localised orbital excitations, and photoionisation cross-sections. & \\
\hline
\end{tabular}
\label{tab4}
\end{table}

\newpage
The usage and importance of the classical and neoclassical algorithms described above can be broken up into the study of the following properties of molecules.
\begin{enumerate}
\item \textbf{Molecular Geometry:} This describes the physical arrangement of atoms in space that constitute a molecule via bond angles and electron pair arrangements. This property determines the intermolecular forces, the polarity, and the molecule's reactivity.
\item \textbf{Magnetic Properties:} This property determines the behaviour of a molecule in the presence of a magnetic field. The spin and orbital motion of the constituent electrons in the atoms create magnetic moments which influence the electronic structure, the symmetry, and the molecular geometry of the molecule.
\item \textbf{Electronic Spectra:} This property facilitates the molecule's absorption and emission of electromagnetic radiation (ER). When the constituent atoms absorb energy, the electrons enter an excited state with higher energy. Conversely, the electrons go into lower energy states when the atoms radiate ER. 
\item \textbf{Chemical Reactivity:} This refers to the ability of the molecule to undergo chemical reactions and is therefore influenced by the presence of chemical functional groups and steric hindrances such as obstructions and overlapping electron cloud repulsions between atoms. 
\end{enumerate}

These properties can be exploited for a plethora of real-world applications. We discuss some of these implementations below.
\begin{enumerate}
\item \textbf{Materials Science and Engineering:} Used for designing novel materials with pertinent properties for a particular application. For predicting the properties of new materials and gaining a deeper understanding of existing materials.
\item \textbf{Nanotechnology:} Used for designing and understanding materials in the nanometre scale ranging from $1-100$ nm. These include nanowires, nanotubes, nanopores, nanocapsules, nanorods, nanofibers, nanopillars, nanostructured membranes, nanocomposites, and dendrimers. 
\item \textbf{Energy Research:} For designing novel energy conversion and storage materials. These include state-of-the-art batteries and solar cells. 
\item \textbf{Physical Chemistry and Chemical Physics:} Perhaps the most ubiquitous application. It is used to study the properties of molecules, the outcomes of chemical reactions, and understand reaction mechanisms. 
\item \textbf{Condensed Matter Physics:} Used to determine materials' magnetic, electronic, and optical properties. 
\item \textbf{Biochemistry and Drug Design:} Used for modelling the structure of DNA, RNA, proteins, biomolecules, and the design of pharmaceutical-grade drugs.
\item \textbf{Environmental Research:} Used to study pollutants, and their impact on ecosystems. Additionally, for the design of green-materials for environmental remediation and pollution-mitigation. 
\end{enumerate}

While these classical methods have been highly successful for decades, and have numerous applications, as discussed above, they possess many drawbacks. We discuss them below.
\begin{enumerate}
\item \textbf{Producing Inaccurate Results:} Since these methods are numerical approximations of the Schr\"{o}dinger equation's description of the electron, errors can easily be carried over and compounded, producing unreliable and imprecise values.
\item \textbf{The Inability to Capture Quantum Mechanical Effects:} These classical methods do not account for Quantum Entanglement or Quantum Tunnelling. Since these phenomena have a bearing on macroscopic physical and chemical properties, this results in coarse-grained results which deviate from experiments in many cases. 
\item \textbf{Inadequacy of Models to be Modular and Transferable:} Computer models / simulations are specific to molecules, and require domain expertise in order to edit code and adjust it for the study of other molecules. In addition, adding new parameters to the model is not a trivial exercise and requires a significant overhaul of the code.
\item \textbf{Limitations and Constraints in the Scope of the Models:} The models are limited to small-electron systems. For larger electron systems, the models become computationally intractable. 
\item \textbf{The Insufficiency in Capturing Theoretical Subtleties and Chemical Accuracy in Reactions:} Classical methods are limited in their Inability to account for abstruse differences in structural dissimilarities, and energy variations. These facets are important in predicting reaction mechanisms.  
\end{enumerate}
Thus, we advocate for, and try to galvanise, the idea of using Quantum Computing (QC) as an alternative method that can be applied as a stand-alone method or in parallel with a classical method. 

The field of QC is rapidly advancing and delving into the potential of quantum mechanics to process information in ways that classical computers are limited in their ability to achieve. Originally introduced by \hyperlink{41}{Feynman, 1982}, who catechised the idea of whether a computer could simulate quantum systems based on the fundamental principles of quantum mechanics. This paved the way for the idea of \textit{qubits} -- the amalgamation of ``quantum'' and ``bits'', that can exist in a superposition of states and can be entangled with one another. Unlike classical bits that can only exist in one of two states, qubits can exist in a continuum of states, providing QC with the ability to perform certain computations faster than classical computers.

One of the most promising QC technologies is based on superconducting circuits that create and manipulate qubits. These circuits operate using microwave signals and are cooled to temperatures close to absolute zero. 

QC has emerged as a promising paradigm for various applications, from Cryptography and Optimisation problems, to Electronic Structure calculations. Various approaches to QC have been explored, including trapped ions, quantum dots, and topological qubits with Quntinuum's recent success in creating non-Abelian anyons -- \textit{nonabelions} (\hyperlink{51}{Iqbal \textit{et al}, 2023}) in the pursuit of fault-tolerant quantum computers being a noteworthy example. However, the realisation of large-scale and dependable quantum computers remains a significant technical challenge that must be overcome to fully exploit QC's potential.

One promising approach to quantum computation is the Variational Quantum Eigensolver (VQE) algorithm, which was introduced by \hyperlink{39}{Peruzzo \textit{et al.}, 2013}. It is a quantum algorithm used to find the lowest eigenvalue of a given Hamiltonian, which corresponds to the ground state energy of a quantum system. VQE is a hybrid algorithm that combines classical and quantum computing resources to determine the lowest eigenvalue of a given Hamiltonian, corresponding to a quantum system's ground state energy. Notably, VQE is designed to run on noisy intermediate scale quantum (NISQ) computers -- see \hyperlink{40}{Preskill, 2018} -- which have a limited number of qubits, and high error rates. The algorithm represents a promising solution for determining the ground state energies of molecules and materials.

The VQE works by computing the expectation value of the Hamiltonian, $\hat{H}$, which is given by Eq. \eqref{eq:vqe}:
\begin{equation}
\label{eq:vqe}
\hat{H}=\sum_{i}\mu_{i}P_{i}, \quad\quad P_{i}\in\left\{X,Y,Z \right\},
\end{equation}
where $\mu_{i}$ are coefficients, and $P_{i}$ are the Pauli matrices $\left\{X,Y,Z\right\}$. Using a Parameterised Quantum Circuit (PQC), $U(\boldsymbol{\theta})$, a trial state, $\ket{\Psi(\boldsymbol{\theta})}$ is prepared as Eq. \eqref{eq:vqe2}:
\begin{equation}
\label{eq:vqe2}
\ket{\Psi(\boldsymbol{\theta})}=U(\boldsymbol{\theta})\ket{0}^{\otimes n}.
\end{equation}
The energy of the system, $E$, given by Eq. \eqref{eq:vqe3}:
\begin{equation}
\label{eq:vqe3}
E(\boldsymbol{\theta})=\bra{\Psi(\boldsymbol{\theta})}\hat{H}\ket{\Psi(\boldsymbol{\theta})}_{\boldsymbol{\theta}},
\end{equation}
is used to determine the minimum energy of the system, $E_{\min}$, given by Eq. \eqref{eq:vqe6}: 
\begin{equation}
\label{eq:vqe6}
E_{\min}=\underset{\boldsymbol{\theta}}{\min}\;E(\boldsymbol{\theta}).    
\end{equation}
The parameters are then optimised using a classical optimiser such as Gradient Descent (GD), Stochastic Gradient Descent (SGD), or any of the other classical optimisers. This iterative process is repeated until a convergence criterion is met, and the final parameters $\boldsymbol{\theta}^*$ are utilised to calculate the ground state energy of the Hamiltonian according to Eq. \eqref{eq:vqe7}:
\begin{equation}
\label{eq:vqe7}
E_{0}=\underset{i}{\min}\;\mu_{i}.
\end{equation}

\begin{figure}[H]
    \centering
    \includegraphics[scale=0.8]{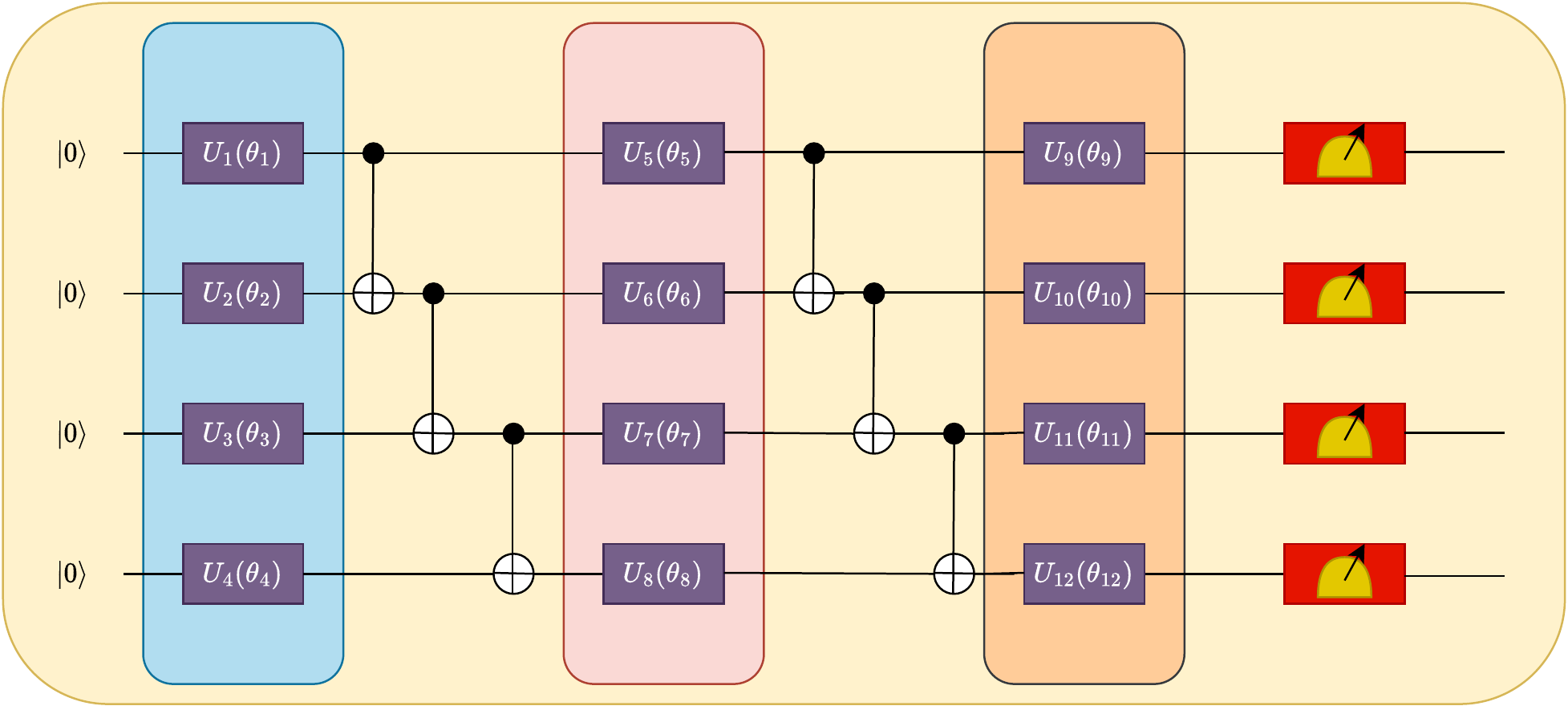}
    \caption{Architecture of the VQE.}
    \label{fig:my_label7}
\end{figure}
As we have discussed the limitations of classical methods used for electronic structure calculations, this research aims to investigate the potential of quantum computing, specifically focusing on the VQE algorithm, to address these challenges and limitations. We believe that the usage of QC for electronic structure calculations offers several advantages over the classical methods; we delineate these below:
\begin{enumerate}
\item \textbf{Accurate Modelling of the Behaviour of the Electrons in the System Under Consideration:}  Since QC is inherently concerned with quantum mechanics, the modelling of quantum mechanical systems in order to predict properties is a more natural and well-furnished method, and can therefore mitigate the inaccuracies of the classical approximation methods.
\item \textbf{Has the Potential to Offer Exponential Speed-up Over Classical Methods:} Since quantum computers have the potential to offer significant speed-ups over classical computers -- prospectively exponential; see for example Grover's algorithm for searching unsorted databases (\hyperlink{46}{Grover, 1996}), and Shor's algorithm used in Cryptography for factoring large numbers and finding discrete logarithms (\hyperlink{45}{Shor, 1997}) -- quantum computers can be used for modelling and simulating larger electron systems.
\item \textbf{Expandability and Scalability of the Algorithms Used to Model and Simulate the Systems Being Studied:} Classical methods become computationally infeasible for large-electron systems, whereas quantum computers have the potential to scale as the molecular systems grow.
\end{enumerate}
Therefore, the applications in this paper are used to show that QC-inspired methods are easier to implement, and produce accurate results within a tolerable margin.

This paper is structured as follows:

\indent In \S\ref{section:my}, we conduct a comprehensive review of prior research to provide a foundation for our study.

\indent In \S\ref{section:my1}, we introduce the new algorithm based on the VQE architecture, and its steps for energy calculation.

\indent In \S\ref{section:my2}, we present our experimental procedures and findings, wherein we compare the performance of VQE against traditional methods such as DFT and HF and demonstrate its proficiency in computing the energies of diverse molecules. 

\indent Finally, in \S\ref{section:my3}, we summarise our findings and elucidate the potential of the VQE to transform the fields of Condensed Matter Physics, and Computational Chemistry.

\section{Literature Review}
\label{section:my}
Several researchers and groups have attempted to calculate atomic properties using QC. Below, we tabulate a non-exhaustive summary of quintessential research papers and provide descriptions of the fruition of their findings (\ref{tab1}, \ref{tab2}, and \ref{tab2-2}).
\begin{table}[H]
\caption{Research papers that have applied QC to electronic structure calculations -- Part I.}
\centering
\begin{tabular}{p{5cm}p{12cm}}
\hline
\hline
\textbf{Literature} &\textbf{Summary} \\
\hline
\hyperlink{16}{Whitfield \textit{et al.}, 2011} &This research highlights the drawbacks of current methods for simulating molecular systems, namely the computational complexity associated with modelling $n$-particle quantum systems. It is shown how pre-computed molecular integrals can be used to obtain the energy of the $n$-particle system using quantum phase estimation, the caveat being that such a system must have a fixed nuclear configuration. For the Hydrogen gas ($\text{H}_{2}$) molecule, the simulation of the chemical Hamiltonian is exhibited on a quantum computer. \\
\hyperlink{13}{Carleo \& Troyer, 2017} &This landmark paper used perceptrons with varying hidden neurons to introduce a variational representation of quantum states for one- and two-dimensional spin models with interaction. The ground state was determined using a Reinforcement Learning-inspired scheme, and the unitary temporal evolution dynamics were described. 

\\
\hline 
\end{tabular}
\label{tab1}
\end{table}
\newpage
\begin{table}[H]
\caption{Research papers that have applied QC to electronic structure calculations -- Part II.}
\centering
\begin{tabular}{p{5cm}p{12cm}} \\
\hline 
\hline
\textbf{Literature} &\textbf{Summary} \\
\hline
\hyperlink{15}{Xia \& Kais, 2018} &In this seminal work, the need for the hybridisation of QC with ML is delineated in order to achieve more accurate results in electronic structure calculations with reduced computational times. A composite model, consisting of a Restricted Boltzmann Machine (RBM), was employed to ascertain the electronic ground state energy for small molecular systems (the demonstrable use case of small molecular systems was chosen because of the current NISQ-era QC technology available). The examples of the Hydrogen ($\text{H}_{2}$), Lithium Hydride (\text{LiH}), and Water ($\text{H}_{2}\text{O}$) molecules were chosen for potential energy surface calculations. \\
\hyperlink{11}{Sureshbabu \textit{et al.}, 2021} &An IBM quantum computer was used to calculate electronic structure on two-dimensional crystal structures -- specifically monolayer hexagonal Boron Nitride ($\text{h}-\text{BN}$) and monolayer Graphene ($\text{h}-\text{C}$) -- using a hybrid method comprising a restricted Boltzmann machine and a quantum machine learning (QML) algorithm. The results were consistent with traditional results from classical methods. \\
\hyperlink{12}{Rossmannek \textit{et al.}, 2021} &An effective Hamiltonian was constructed by incorporating a mean-field potential on a restricted Action Space (AS) via an embedding scheme. Using the VQE algorithm, the ground state of the AS Hamiltonian, $\hat{H}_{0}$, was calculated. \\
\hyperlink{14}{Song \textit{et al.}, 2023} &Using VQE circuits on Quantinuum's ion-trap quantum computer H1-1, small molecules were simulated with plane-wave basis sets. Using a small number of iterations, the results from Correlation Optimised Virtual Orbitals (COVO) were replicated within the tolerable range of $11\;\text{mE}_{\text{h}}$ (milli-Hartree) $\approx 0.2993\;\text{eV}$.  \\

\hline 
\end{tabular} 
\label{tab2}
\end{table}

\begin{table}[H]
\caption{Research papers that have applied QC to electronic structure calculations -- Part III.}
\centering
\begin{tabular}{p{5cm}p{12cm}} \\
\hline 
\hline
\textbf{Literature} &\textbf{Summary} \\
\hline
\hyperlink{17}{Naeij \textit{et al.}, 2023} &This study used a VQE to calculate the ground state energy of Protonated Molecular Hydrogen ($\text{H}_{3}^{+}$), Hydroxide ($\text{OH}^{-}$), Hydrogen Fluoride ($\text{HF}$), and Borane ($\text{BH}_{3}$). The Unitary Coupled Cluster for Single and Double excitations (UCCSD) is used to construct an ansatz with the fermion-to-qubit and parity transformation. Lastly, this hybrid VQE method was benchmarked against the Unrestricted Hartree-Fock (UHF) and FCI classical methods, and high fidelity between the classical and quantum approaches is shown. \\
\hyperlink{70}{Qing \& Xie, 2023} &  This study implements the VQE using Qiskit / IBM Quantum to determine the ground state energy of a Hydrogen (H) molecule, the results reveal that VQE is highly effective in accurately computing molecular properties. Nevertheless, the study also highlights certain challenges and constraints in  scaling the algorithm for larger molecules.  
\\
\hline 
\end{tabular} 
\label{tab2-2}
\end{table}
\section{Theory}
\label{section:my1}
In order to perform electronic structure calculations of molecules on a quantum computer, we require the following steps:

\begin{enumerate}
\item Write down the Born-Oppenheimer Hamiltonian, $\hat{H}$, for the molecule in terms of creation, $a_{i}^{\dagger}$, and annihilation, $a_{i}$, operators. 
\item Convert this Hamiltonian into matrix form by applying a suitable fermionic transformation that take the creation and annihilation operators into single-bit quantum gates. There are several methods of doing this, and Tab. \ref{tab5} outlines these techniques. We denote the Pauli spin matrices as $X, Y, Z$ for the $x, y, z$ unitary evolutions, respectively, and $I$ for the $2\times 2$ identity matrix.

\item Solve for the ground state, and excited states of the molecular system, The current state-of-the-art is the VQE method, which we will adopt in all subsequent calculations. Notwithstanding, the Phase Estimation Algorithm (PEA)  -- see \hyperlink{50}{Kitaev, 1995}; \hyperlink{51}{Kitaev, 1997} -- is also widely used.

\end{enumerate}
\begin{table}[H]
\caption{Various mappings of creation, $a_{i}^{\dagger}$, and annihilation, $a_{i}$, operators to single-bit quantum gates.}
\centering
\begin{tabular}{p{6cm}p{11cm}}
\hline
\hline
\textbf{Method} &\textbf{Mapping} \\
\hline
Jordan-Wigner transformation \newline (\hyperlink{35}{Jordan \& Wigner, 1928}; \hyperlink{16}{Whitfield \textit{et al.}, 2011}; \hyperlink{34}{Fradkin, 1989}) & \footnotesize{$a_{i}^{\dagger}\longrightarrow\frac{1}{2}\{\left(X_{i}-\imath Y_{i}\right)\otimes_{j=1}^{i-1}Z_{j},$}  \newline
 \footnotesize{$a_{i}\longrightarrow\frac{1}{2}\left(X_{i}+\imath Y_{i}\right)\otimes_{j=1}^{i-1}Z_{j}.$} \newline
 For a $2n$-electron system, the computational complexity of this method is $\mathcal{O}(n)$. \\
\hline
Binary code transformation \newline (\hyperlink{36}{Steudtner \& Wehner, 2018}) &\footnotesize{$a_{i}^{\dagger}\longrightarrow\frac{1}{2}\left\{X^{U(i)}\otimes\left[1+Z^{F(i)}\right]\otimes Z^{\mathcal{P}(i)}\right\},$} \newline
\footnotesize{$a_{i}\longrightarrow\frac{1}{2}\left\{X^{U(i)}\otimes\left[1-Z^{F(i)}\right]\otimes Z^{\mathcal{P}(i)}\right\},$} \newline
where $U(i)$ is the qubit update set when the creation and annihilation operators are applied to orbital $i$, $F(i)$ is a checking function that ascertains whether the creation and annihilation operators yield $0$, and $\mathcal{P}(i)$ is a parity-check function that checks the phase change when the creation and annihilation operators are applied to spin orbital $i$. The computational complexity will be dependent on the value of $U(i), F(i),$ and $\mathcal{P}(i)$. 
 \\
Parity transformation \newline (\hyperlink{37}{Seeley \textit{et al.}, 2012}) &\footnotesize{$a_{i}^{\dagger}\longrightarrow\frac{1}{2}\left(\otimes_{j=1+1}^{n}X_{j}\otimes X_{i}\otimes Z_{i-1}-\imath\otimes_{j=1+1}^{n}X_{j}\otimes Y_{i}\right)$}, \newline
\footnotesize{$a_{i}\longrightarrow\frac{1}{2}\left(\otimes_{j=1+1}^{n}X_{j}\otimes X_{i}\otimes Z_{i-1}+\imath\otimes_{j=1+1}^{n}X_{j}\otimes Y_{i}\right)$}, \newline
for a $2n$-electron system. The computational complexity of this method is $\mathcal{O}(n)$. \\
\hline
Bravyi-Kitaev transformation \newline (\hyperlink{37}{Seeley \textit{et al.}, 2012}; \hyperlink{38}{Tranter \textit{et al.}, 2015}) &\footnotesize{$a_{i}^{\dagger}\longrightarrow\frac{1}{2}\left[X^{U(i)}\otimes X_{i}\otimes Z^{\mathcal{P}(i)}-\imath X^{U(i)}\otimes Y_{i}\otimes Z^{\mathcal{P}(i)}\right],$} \newline
\footnotesize{$a_{i}\longrightarrow\frac{1}{2}\left[X^{U(i)}\otimes X_{i}\otimes Z^{\mathcal{P}(i)}+\imath X^{U(i)}\otimes Y_{i}\otimes Z^{\mathcal{P}(i)}\right]$}, \newline
where $U(i)$ and $\mathcal{P}(i)$ have their regular meanings. This method combines the Jordan-Wigner and Parity transformations, and for a $2n$-electron system, it has a quasilinear computational complexity $\mathcal{O}(n\log n)$. \\
\hline
\end{tabular}
\label{tab5}
\end{table}

Below we computationalise the calculational steps for electronic structure calculations of molecular systems using quantum computers in the form of the algorithm \ref{QElectra} below.

\begin{algorithm}[H]
\caption{\texttt{QElectra$(\Psi_{0})$}}
\begin{algorithmic}
\State \texttt{// This algorithm takes in a ground state, $\Psi_{0}$, and returns the ground and excited state energies} 
\State \textbf{input} basis functions $\chi_{i}, \chi_{j}, \chi_{k}, \chi_{l}$; creation operators $a_{i}^{\dagger}, a_{j}^{\dagger}$, annihilation operators $a_{j}, a_{k}, a_{l}$; tolerance $\mathscr{E}$; perturbation factor $\lambda$
\State \textbf{initialise} $E_{0}$ \texttt{// initialising the ground state energy $E_{0}$}
\For{each atom in the molecule} 
  \State calculate the nuclear repulsion / Coulomb energy
  \begin{equation*}
h_{0}=\frac{1}{4\pi\varepsilon_{0}}\sum_{i=1}^{N_{\text{atom}}}\sum_{j=i+1}^{N_{\text{atom}}}\frac{z_{i}z_{j}}{|\mathbf{r}_{i}-\mathbf{r}_{j}|}
  \end{equation*}
  \State calculate the one-electron operator
  \begin{equation*}
\hat{\mathscr{F}}=-\frac{1}{2}\sum_{i}\frac{1}{\omega_{i}}\nabla^{2}_{i}-\sum_{A, i}\frac{\mathcal{Z}_{A}}{r_{iA}}+\sum_{i<j}\frac{1}{r_{ij}}
  \end{equation*}
  \State calculate the one-electron orbital integral
  \begin{equation*}
h_{ij}=\bra{\chi_{i}}\hat{\mathscr{F}}\ket{\chi_{j}}=\int_{-\infty}^{+\infty}\chi_{i}^{*}(\mathbf{r})\hat{\mathscr{F}}(\mathbf{r})\chi_{j}(\mathbf{r})\;d\mathbf{r}
  \end{equation*}
  \State calculate the two-electron orbital integral
  \begin{equation*}
h_{ijkl}=\int_{-\infty}^{+\infty}\int_{-\infty}^{+\infty}\frac{\chi_{i}(\mathbf{r}_{1})\chi_{j}(\mathbf{r}_{1})\chi_{k}(\mathbf{r}_{2})\chi_{l}(\mathbf{r}_{2})}{r_{12}}\;d\mathbf{r}_{1}d\mathbf{r}_{2}
  \end{equation*}
  \State calculate the second quantisation Hamiltonian operator
  \begin{equation*}
\hat{H}=h_{0}+\sum_{i, j}h_{ij}a_{i}^{\dagger}a_{j}+\sum_{i, j, k, l}h_{ijkl}a_{i}^{\dagger}a_{j}^{\dagger}a_{k}a_{l}
  \end{equation*} 
\EndFor
\State map the Hamiltonian operator $\hat{H}$ to single-bit quantum gates using a suitable transformation as described in Tab. \ref{tab5} 

\algstore{testcont}
\end{algorithmic}
\label{QElectra}
\end{algorithm}
\newpage
\begin{algorithm}[H]
\ContinuedFloat
\caption{\texttt{QElectra }$(\Psi_{0})$ -- Part 2}
\begin{algorithmic}
\algrestore{testcont}
\While{$||E_{0}^{i+1}-E_{0}^{i}||>\mathscr{E}$}
  \State calculate the ground state energy, and excited states using the VQE ansatz 
  \begin{equation*}
E_{0}=\frac{\bra{\Psi_{0}(\boldsymbol{\theta})}\hat{H}\ket{\Psi_{0}(\boldsymbol{\theta})}}{\bra{\Psi_{0}(\boldsymbol{\theta})\ket{\Psi_{0}(\boldsymbol{\theta})}}}, \quad\quad E_{k}=E_{0}+k\lambda
\end{equation*}

  \State update $\boldsymbol{\theta}_{i+1}\longleftarrow\boldsymbol{\theta}_{i}$ according to the chosen optimisation method (GD, SGD, Adam, etc.)
\EndWhile
\State \textbf{return} $E_{0}, E_{k}$
\end{algorithmic}
\label{QElectra2}
\end{algorithm}
In algorithm \ref{QElectra} above, the first step entails calculating the nuclear binding energy $h_{0}$. The variables in its associated formula are: $\varepsilon_{0}$, the permittivity of free space; $N_{\text{atom}}$, the number of atoms in the molecular system being studied; $z_{i}, z_{j}$, the atomic numbers of the $i^{\text{th}}$ and $j^{\text{th}}$ atoms respectively; and $\mathbf{r}_{i}$ and $\mathbf{r}_{j}$, the position vectors of the $i^{\text{th}}$ and $j^{\text{th}}$ atoms respectively, in reference to a defined coordinate system.

Secondly, we calculate the one-electron operator, $\hat{\mathscr{F}}$. The variables associated with the formula are: $\omega_{i}=\kappa m_{\text{e}^{-}}$, the effective mass of the $i^{\text{th}}$ electron (the rest mass of the electron, $m_{e^{-}}$, scaled by some factor $\kappa$); $\mathcal{Z}$, the charge of the $A^{\text{th}}$ nucleus; $r_{iA}$, the distance between the $i^{\text{th}}$ and $A^{\text{th}}$ nucleus; and $r_{ij}$, the distance between the $i^{\text{th}}$ and $j^{\text{th}}$ nucleus. 

Thirdly, we calculate the one-electron orbital integral, $h_{ij}$. The variables associated with the formula are: $\chi_{i}$ and $\chi_{j}$, the $i^{\text{th}}$ and $j^{\text{th}}$ basis functions respectively. 

In the fourth step, we calculate the two-electron integral, $h_{ijkl}$, and the variables associated with this formula have the same meaning as the variables above.

In the sixth step, we use the previously calculated variables in order to compute the Hamiltonian operator for the system, $\hat{H}$.

Lastly, whilst the successive ground state energies are greater than some tolerance threshold $\mathscr{E}\simeq 0$ in each iteration, the VQE approximation method is used to calculate the ground state energy $E_{0}$, and the energy of the $k^{\text{th}}$ excited state, $E_{k}$, for some perturbation factor $\lambda\in\left[0,1\right]$, with $\lambda\simeq 0.1$ for small perturbations, and $\lambda\rightsquigarrow 1$ for large perturbations.  

Finally, once the stopping criterion is met, i.e. when the difference between successive ground state energies is smaller than the tolerance, the algorithm returns the ground state and $k^{\text{th}}$ excited state energies. 

\begin{tcolorbox}
$^{\ddagger}$We unequivocally point out the caveat that while we do not claim to introduce a revolutionary, paradigm-shifting algorithm that is groundbreaking in the form of \ref{QElectra}, the original contribution, and saliency of the method championed, in this paper is the consolidation of, and systemization of ideas and methodologies adopted by researchers working in the field, who use hybrid Classical-Quantum (CQ) approaches to perform electronic structure calculations -- as seen in the literature cited -- into one synthesised form, which is comprehensive. We would also like to mention the modularity property of Algorithm \ref{QElectra}: We have used the VQE approach to find the minimum energy, however, we have seen in the literature that some researchers have used the PEA.  
\end{tcolorbox}

\section{Experiments, Results, and Discussion}
\label{section:my2}
This study aimed to determine the most precise and efficient method for computing electronic energies through a comparative analysis of three approaches: HF, DFT, and VQE. The investigation focused on assessing the efficacy of QC in this process, aiming to demonstrate the superiority of quantum methods over classical approaches. The efficacy of several techniques were evaluated through the computation of energy values for diverse molecules, including Water ($\text{H}_{2}\text{O}$), Lithium Hydride ($\text{LiH}$), Methane ($\text{CH}_4$), Ammonia ($\text{NH}_3$), and Carbon Dioxide ($\text{CO}_2$).

For the DFT calculations, we generated the electronic density of the system using a selected exchange-correlation functional and initial atomic coordinates. We then solved the Kohn-Sham equations (\ref{eq:kohnsham}), yielding the electronic wavefunctions and energies. This was accomplished through iterative Self-Consistent Field (SCF) calculations, repeated until convergence. 

Similarly, in the HF calculations, the processes was initiated with an initial guess for the wavefunction. The Hartree-Fock equations (\ref{eq:hf}) were subsequently solved in order to obtain the self-consistent wavefunction and energies, and again utilised iterative SCF calculations until convergence was achieved. The total energy was calculated for both cases, and convergence of the electronic wavefunctions and energies was accomplished. The results from the QC approach were then compared with the results obtained through the DFT and HF methods.

For the VQE calculations, firstly the electronic structure of each molecule was obtained, and the electronic Hamiltonian was successively mapped to a qubit Hamiltonian using the parity mapper. Using the qubit converter, the  qubit Hamiltonian was converted to the Pauli basis, expressed in terms of fermionic operators, to a qubit Hamiltonian exhibited in terms of Pauli operators. This conversion is necessary because quantum computers typically operate with qubits. Using the Jordan-Wigner transformation method, the qubit converter maps the fermionic problem to the qubit problem. The result is a qubit Hamiltonian expressed as a linear combination of Pauli operators. the Unitary Coupled Cluster (UCC) ansatz was deployed with single and double excitations to construct a trial wavefunction for the VQE computation. 

This ansatz involves representing the wavefunction as a linear combination of exponentially parameterised unitary operators acting on a reference state, which can be expressed as:
$|\Psi_{\text{UCC}}\rangle = e^{T} \ket{\Phi}$,
where $T$ is a cluster operator that generates the single and double excitations from the reference state $|\Phi\rangle$. In order to carry out the VQE calculation, the Simultaneous Perturbation Stochastic Approximation (SPSA) optimiser was utilised with a fixed number of iterations; this optimiser estimates the gradient of the cost function using two randomly perturbed function evaluations, and updates the parameters of the ansatz in the direction that minimises the estimated gradient, as given by the following update rule:
$\theta_{k} \longleftarrow \theta_{k} - a_k \tilde{\nabla}_{\theta_{k}} J(\theta_k)$,
where $\theta_{k}$ and $\theta_{k+1}$ are the parameter vectors at iteration $k$ and $k+1$, $a_k$ is the step size, $\tilde{\nabla}_{\theta_{k}} J(\theta_k)$ is the estimated gradient of the cost function at $\theta_k$, and $J(\theta_k)$ is the cost function itself. The VQE calculations were performed on the AER simulator backend provided by \hyperlink{60}{IBM}.

\begin{figure}[!ht]
    \centering
    \includegraphics[scale=0.5]{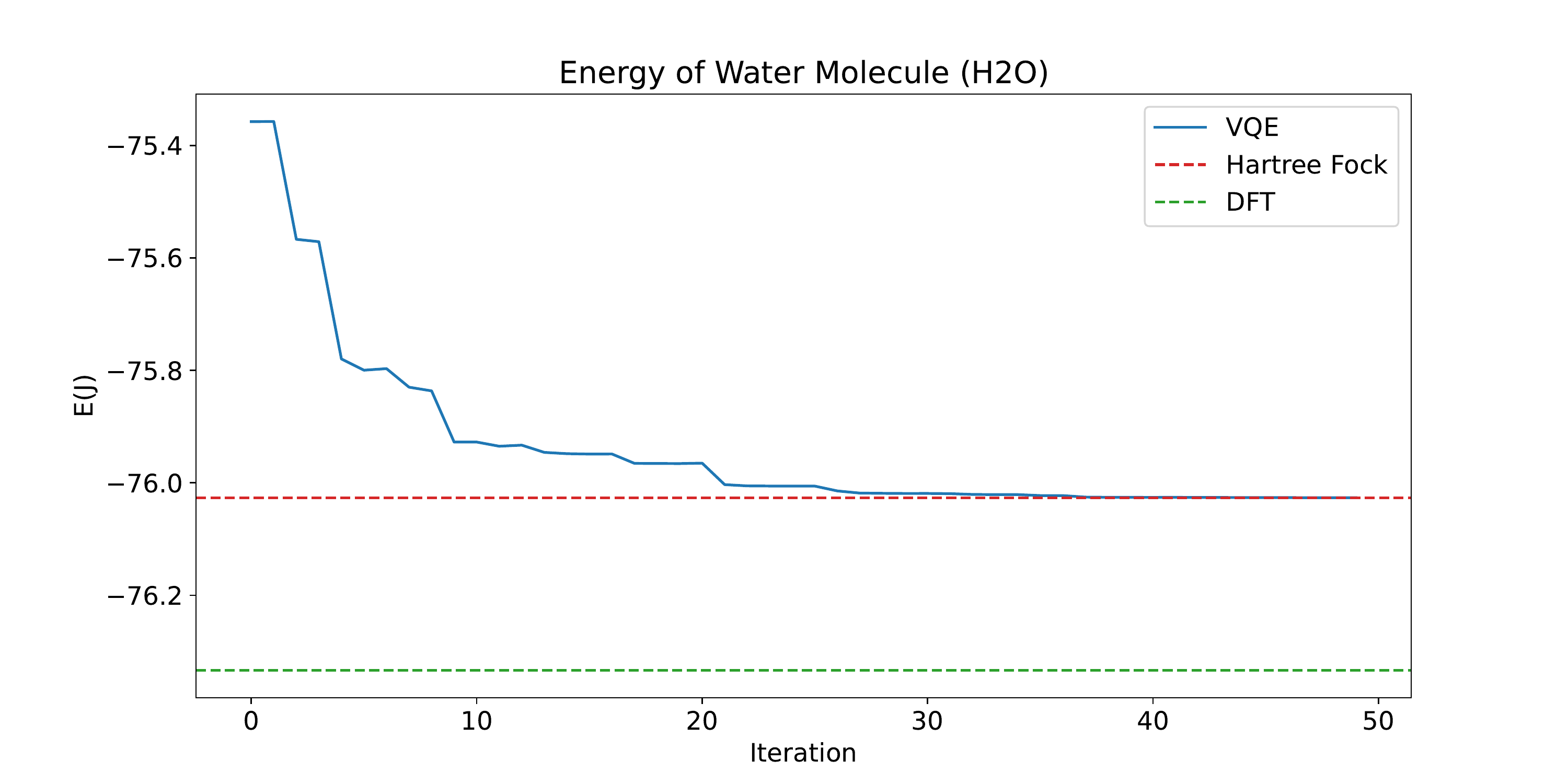}
    \caption{Electronic structure calculations for Water molecule $(H_{2}O)$: Method comparison.} 
    \label{fig:h2o}
\end{figure}

\begin{figure}[!ht]
    \centering
    \includegraphics[scale=0.5]{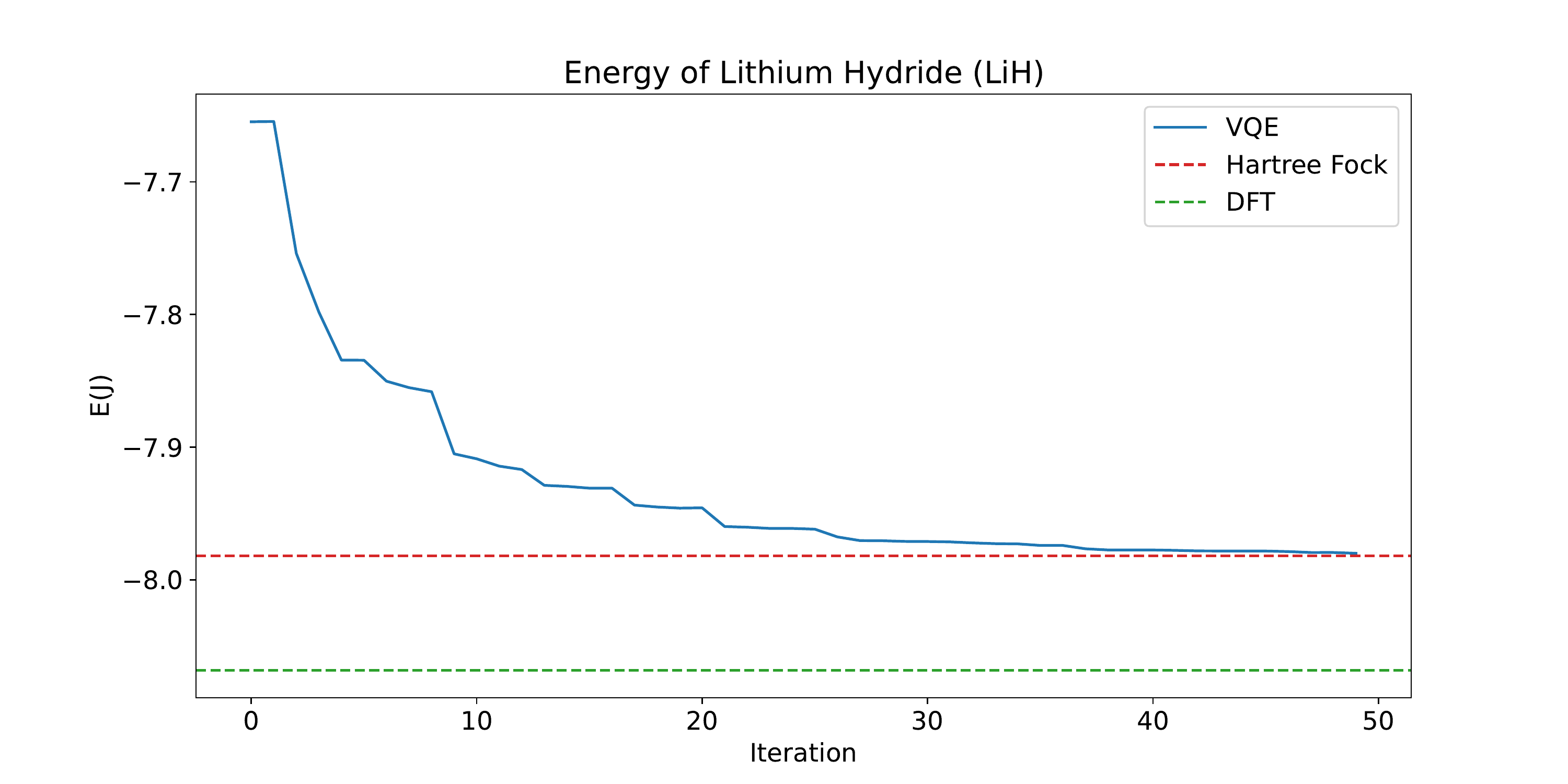}
    \caption{Electronic structure calculations for Lithium Hydride molecule (LiH): Method comparison.} 
    \label{fig:lih}
\end{figure}

\begin{figure}[!ht]
    \centering
    \includegraphics[scale=0.5]{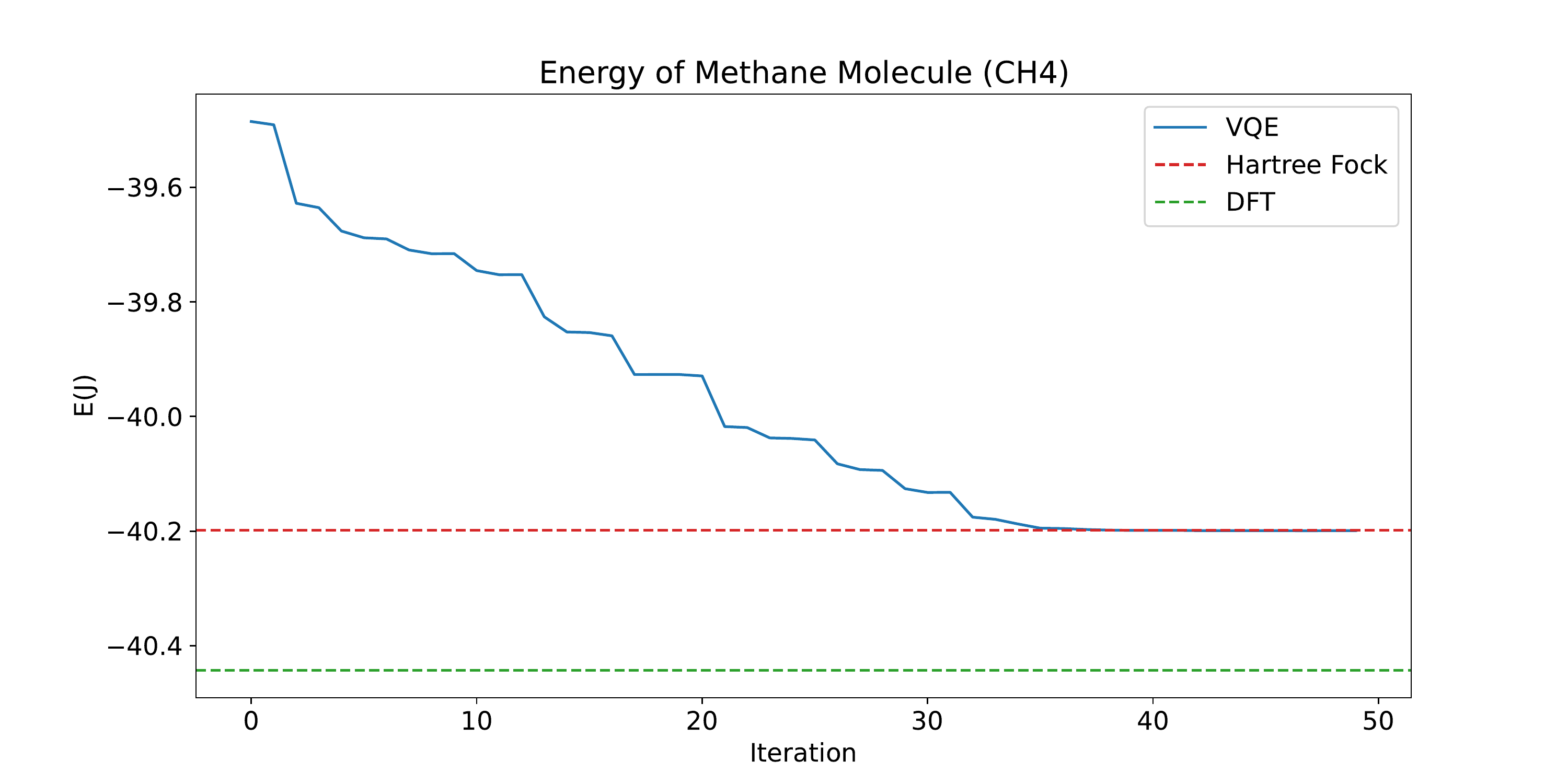}
    \caption{Electronic structure calculations for Methane molecule $(CH_4)$: Method comparison.} 
    \label{fig:ch4}
\end{figure}

\begin{figure}[!ht]       
    \centering
    \includegraphics[scale=0.5]{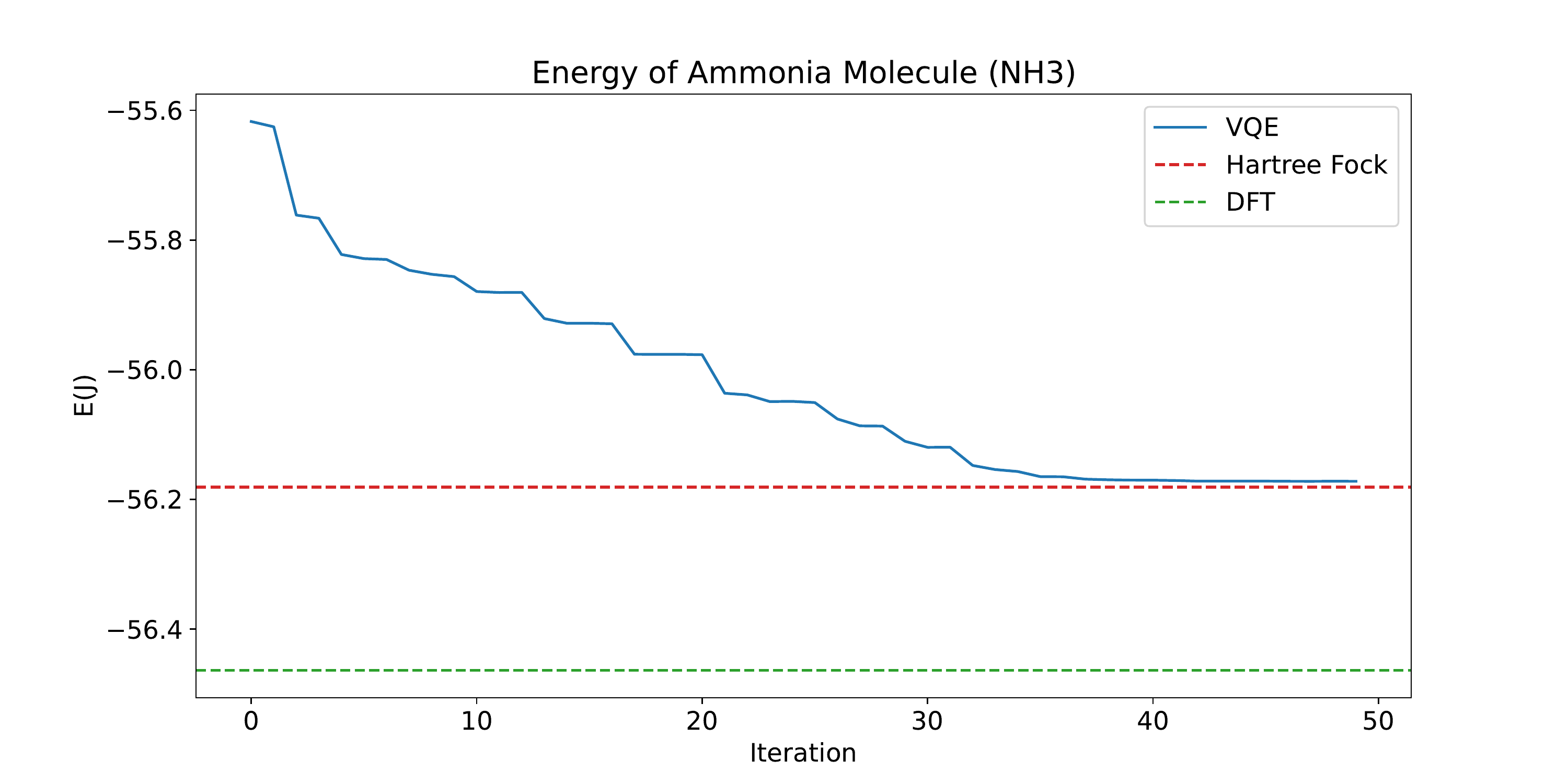}
    \caption{Electronic structure calculations for Ammonia molecule (NH3): Method comparison.} 
    \label{fig:nh3}
\end{figure}

\begin{figure}[!ht]
    \centering
    \includegraphics[scale=0.5]{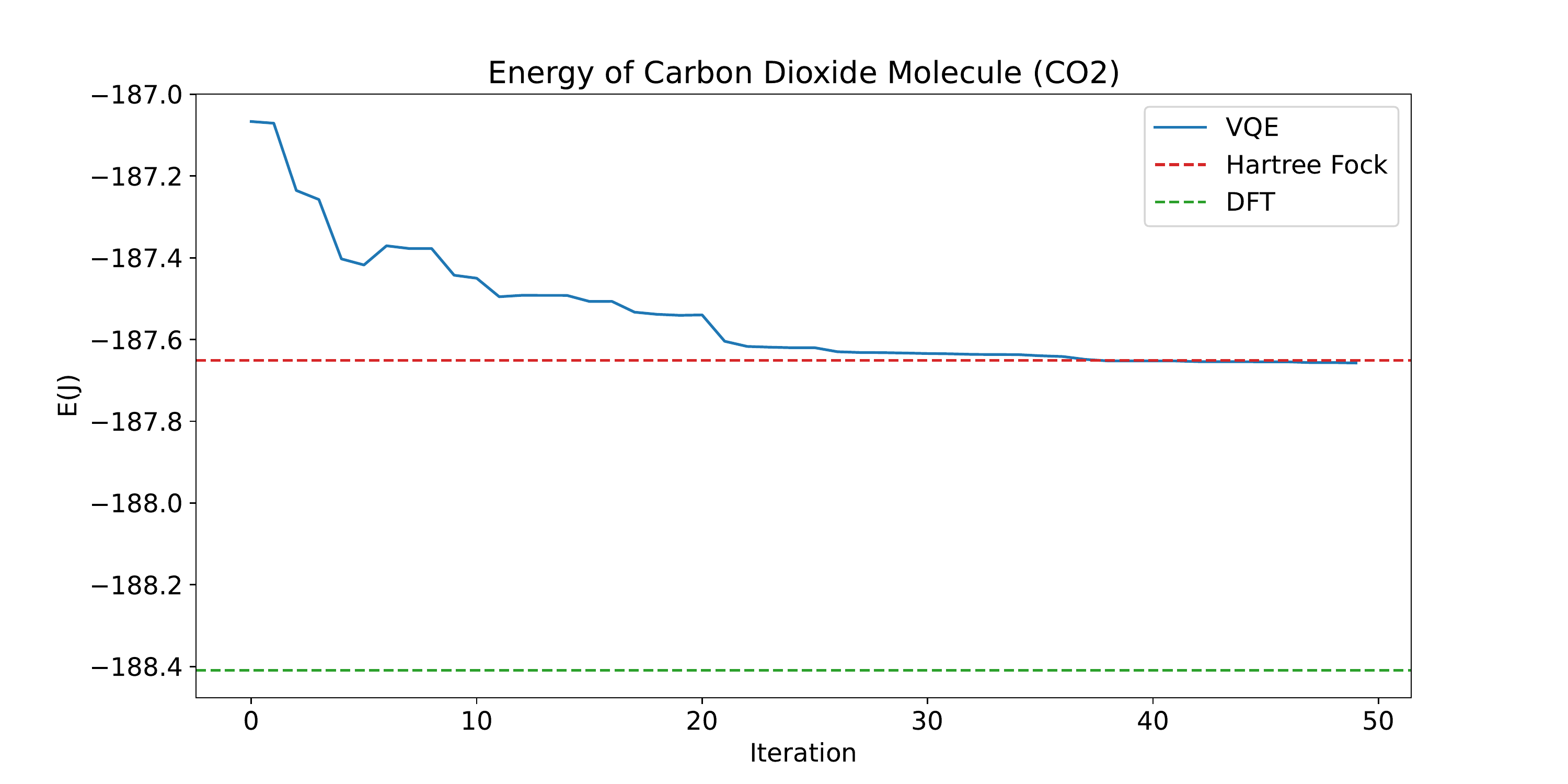}
    \caption{Electronic structure calculations for Carbon Dioxide molecule $(CO_{2})$: Method comparison.} 
    \label{fig:co2}
\end{figure}

The results obtained demonstrate that the VQE method consistently outperformed the HF and DFT methods in terms of both accuracy and efficiency. For example, in the case of Water, the energy values computed using the HF, DFT, and VQE methods were $-76.02679364497443\;\text{J}$, $-76.33340861478466\;\text{J}$, and $-76.02657123746106\;\text{J}$, respectively, as shown in Fig. \ref{fig:h2o}. The VQE energy value agrees with the HF value, and is much more accurate than the DFT value. This trend is consistent across all the molecules studied, as shown in Figures \ref{fig:lih}, \ref{fig:ch4}, \ref{fig:nh3}, and \ref{fig:co2}, indicates that the VQE is a robust and reliable method for computing electronic energies. Based on the graphs and Tab. \ref{tab:energies}, which summarised the energy values obtained using the HF, DFT, and VQE methods for each molecule, these results demonstrate that the VQE method can potentially provide a more efficient and accurate approach to determining the energy of electronic structures, especially for complex molecules where DFT may not provide accurate results.

Notably, the DFT energy values are consistently lower than the HF and VQE values. This can be attributed to the inherent approximations used in the DFT method, which may not accurately capture the exact behaviour of the electrons in the system.
These findings have important implications for Computational Chemistry, as we have demonstrated that QC can provide a more efficient and accurate method for computing electronic energies than classical approaches. 
\newpage

\newpage
\newpage
\begin{table}[h]
    \centering
    \caption{Energy values obtained using the HF, DFT, and VQE methods for each molecule.}
    \label{tab:energies}
    \begin{tabular}{|c|c|c|c|}
        \hline
        \multirow{2}{*}{Molecule} & \multicolumn{3}{c|}{Method} \\ \cline{2-4} 
        & HF & DFT & VQE \\ \hline
        Water ($\text{H}_{2}\text{O}$) &  $-76.02679364497443\;$J & $-76.33340861478466\;$J & $-76.02657123746106\;$J \\ \hline
        Lithium Hydride (LiH) & $-7.981767664359352\;$J & $-8.068192292902214\;$J & $-7.979985984912321\;$J \\ \hline
        Methane ($\text{CH}_{4}$) & $-40.19870325538812\;$J & $-40.44299420579781\;$J & $-40.19911992417514\;$J \\ \hline
        Ammonia ($\text{NH}_{3}$) & $-56.18109675851954\;$J & $-56.46351100537343\;$J & $-56.172108720433144\;$J \\ \hline
        Carbon Dioxide ($\text{CO}_{2}$) & $-187.65110770987644\;$J & $-188.4094301538952\;$J & $-187.6573437805891\;$J \\ \hline
    \end{tabular}
\end{table}

\section{Conclusion}
\label{section:my3}
In summary, this research paper has delved into the promising potential of QC, with specific emphasis on the usage of the VQE algorithm, for electronic structure calculations. This study has identified the limitations and complexities by thoroughly comparing traditional electronic structure calculation methods, including HF theory, DFT, and CC. Our findings indicate that the VQE algorithm provides a more efficient solution to these limitations due to quantum parallelism. The theory section of this paper has elaborated on the VQE algorithm in detail, including the creation and annihilation of operator mappings to single-bit quantum gates. 

This study has also showcased the power of the VQE by calculating the energies of five different molecules, and comparing values obtained from traditional methods. Our research indicates that the VQE can achieve similar energy values using fewer computational resources. The implications of our research affirms that QC has the potential to revolutionise the field of Computational Chemistry, providing a new paradigm in electronic structure calculations with wide-ranging applications in Materials Science, and Physics. Furthermore, this study highlights the need for continued development of QC algorithms and hardware to fully realise this technology's potential.

In conclusion, this research paper provides a comprehensive introduction to electronic structure calculations and a particular QC approach to these calculations, highlighting the potential of the VQE as a robust algorithm in this domain. The results of this study have consequential implications for the field of Computational Chemistry, and demonstrate the transformative potential of QC in this area. Lastly, this study sets the stage for future research and development in this exciting field.

\end{document}